\documentclass[conference]{IEEEtran}

\usepackage{cite}      

\usepackage{graphicx}  

\usepackage{subfigure} 

\usepackage{amsmath}   

\newcommand{\eVdist}{\kern-0.06667em}

\newcommand{\tev}{{\,\mathrm{Te}\eVdist\mathrm{V\/}}}

\newcommand{\gev}{{\,\mathrm{Ge}\eVdist\mathrm{V\/}}}
\newcommand{\met}{{\,\mathrm{m}}}

\newcommand{\km}{{\,\mathrm{km}}}

\newcommand{\khz}{{\,\mathrm{kHz}}}

\begin{document}

\title{KM3NeT: A Next Generation Neutrino Telescope\\ in the Mediterranean Sea}

\author{\authorblockN{Alexander Kappes}
\authorblockA{University Erlangen-Nuremberg\\
Physics Institute\\
D-91058, Erlangen, Germany\\
Email: kappes@physik.uni-erlangen.de}\\
for the KM3NeT Consortium
}



\maketitle

\begin{abstract}
To complement the IceCube neutrino telescope currently under
construction at the South Pole, the three Mediterranean neutrino
telescope projects ANTARES, NEMO and NESTOR have joined forces to
develop, construct and operate a km$^3$-scale neutrino telescope in
the Mediterranean Sea. Since February 2006, the technical
specifications and performance of such a detector are studied in the
framework of a 3-year EU-funded Design Study. In 2009 a technical
design report will be released laying the foundations for the
construction of the detector. In the following, the current status of
the Design Study is presented and examples of solutions for the
technical challenges are discussed.
\end{abstract}


%
\IEEEpeerreviewmaketitle

\section{Introduction}
The detection of high energy neutrinos from astrophysical sources
would be a major breakthrough in our understanding of origin and
production mechanisms of cosmic rays. Unfortunately, despite intense
search for these neutrinos during recent years no such neutrino has
been identified up to now. Recent calculations
\cite{apj:656:870,pr:d74:063007} indicate that detectors with at least
a km$^3$ of instrumented volume are required for this task where
detectors of the first generation like AMANDA or ANTARES have typical
volumes of 0.01\,km$^3$.

At the South Pole the IceCube detector with an instrumented volume of
1\,km$^3$ is currently being build as the successor to the AMANDA
neutrino telescope. However, due to the large atmospheric muon
background for upward observations with neutrino telescopes, the
sensitivity of the detector to sources in the southern sky which
includes most of the Galactic Plane and the Galactic Centre is greatly
reduced. These regions harbour many potential high energy neutrino
sources like supernova remnants, pulsar wind nebulae, microquasars and
other binary systems, but also unidentified sites of high energy
gamma-ray emissions. In order to be able to observe these sources, a
km$^3$-scale neutrino telescope in the Northern Hemisphere is
required.

Building on the experience gained in the pilot projects ANTARES, NEMO
and NESTOR, the three collaborations have joined forces to develop,
construct and operate such a km$^3$-scale neutrino telescope, KM3NeT,
in the Mediterranean Sea at the beginning of the next decade. At the
time of this article the consortium consists of 37 institutes from 10
European countries (Cyprus, France, Germany, Greece, Ireland, Italy,
Netherlands, Romania, Spain, UK). Further, also non-European
institutions are welcome to join the collaboration. KM3NeT is
envisioned as a multidisciplinary research infrastructure with a
permanent deep-sea access for marine sciences (such as oceanology,
marine biology, environmental sciences, geology and geophysics)
through an \emph{associated science node}.

\section{Status of the KM3NeT project}
Evaluating the experience gained with the pilot projects it became
clear that a simple scale-up of these detectors to km$^3$ size is
technically not feasible and/or too expensive. Therefore, a research
and development phase was initiated which is conducted within the
framework of a Design Study funded by the EU in FP6 with 9 MEuro
(total volume ca. 20 MEuro). It started in February 2006 and will end
in 2009. The main goal of the Design Study is the compilation of a
technical design report (TDR) that subsequently allows for a timely
construction of the detector and its concurrent operation with
IceCube. The overall cost for the KM3NeT infrastructure is estimated
to lie between 220 and 250 MEuro.

KM3NeT is recognised by ESFRI (European Strategy Forum on Research
Infrastructures) as \emph{a research infrastructure of pan-European
interest} and is listed on the ESFRI roadmap \cite{web:esfri:roadmap}
for future large scale infrastructures. This entitled the consortium
to be funded in the framework of a \emph{Preparatory Phase} in the EU
FP7 program which will address the political, financial, governance,
strategic and remaining technical issues. This process will also lead
to a decision concerning the choice of the site for the construction
of KM3NeT. Apart from environmental parameters relevant to the physics
performance of the detector this also involves socio-political and
regional considerations.

\section{Detector performance studies}
The aim of the Design Study is to deliver the design specifications
for a detector which yields the best physics performance for a given
budget. Therefore, in the first phase of the Design Study a large
number of basic detector configurations have been simulated and their
performance with respect to astrophysical benchmark fluxes has been
evaluated. The minimal performance objectives are an effective volume
of $1 \km^3$ with an angular resolution of $0.1^\circ$ for muons with
energies above $10\tev$. The detector has to be sensitive to all
neutrino flavors and the lower energy threshold should be at a few
$100 \gev$ (about $100 \gev$ for selected targets).

The basic building block of a detector is the optical module (OM)
which contains the photomultiplier(s) (PMTs) for the detection of the
Cherenkov light from charged particles. Up to now, neutrino telescopes
have been using a single (typically 10'') PMT mounted in a pressure
resistant glass sphere. In the case of ANTARES three of these optical
modules are mounted on one frame, looking downward at an angle of
$45^\circ$. AMANDA and IceCube are using a single large PMT per OM,
facing downward.

\begin{figure}
\centering
\includegraphics[width=2.in]{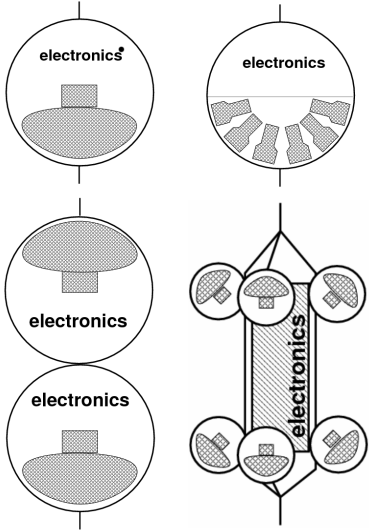}
\caption{Examples of optical module designs that were evaluated in
different detector layouts with respect to their performance. The two
OMs on the left side and the one in the lower right corner contain
10'' PMT(s) whereas the one in the upper right corner is build up of
21 3'' PMTs (taken from \cite{thesis:kuch:2007}).}
\label{fig:km3net-PMT_layouts}
\end{figure}

In the course of a detailed simulation \cite{thesis:kuch:2007}, a
large variety of possible OM configurations was studied (examples can
be seen in Fig.~\ref{fig:km3net-PMT_layouts}), among others a
configuration that uses several small 3'' PMTs either arranged in a
half sphere (Fig.~\ref{fig:km3net-PMT_layouts} upper right) or in a
full sphere (not shown).
Small PMTs have a higher quantum efficiency, a better
single photon resolution and a smaller transit time spread. Also, the
usage of several PMTs in an OM further improves the single photon
counting capability and yields directional sensitivity which helps in
suppressing the optical background from bioluminescence. 

\begin{figure*}
\centerline{\subfigure[Cuboid]{\includegraphics[height=2.in,angle=-90]{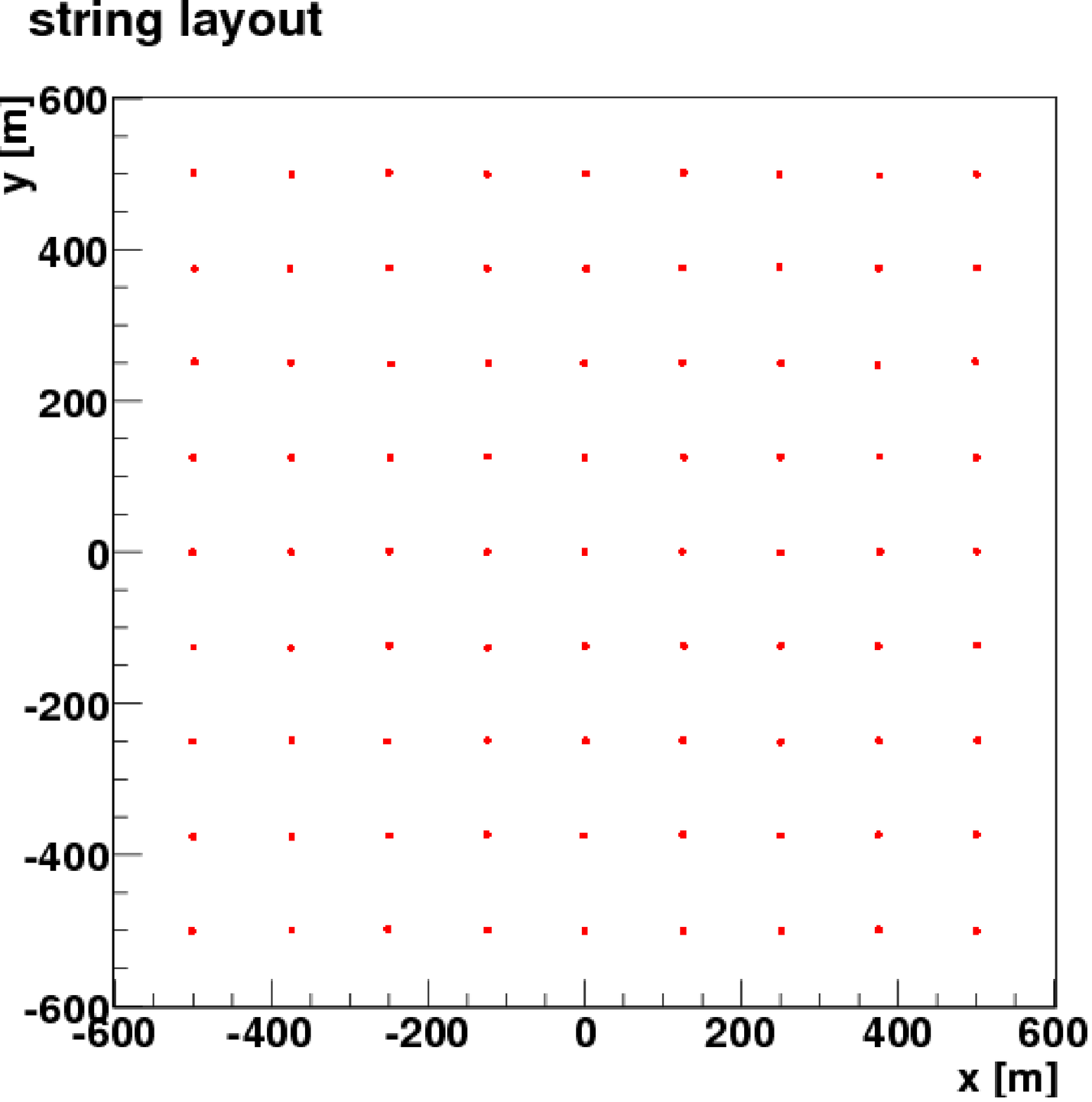}
\label{fig:km3net-cuboid}}
\hfil
\subfigure[Ring]{\includegraphics[height=2.in,angle=-90]{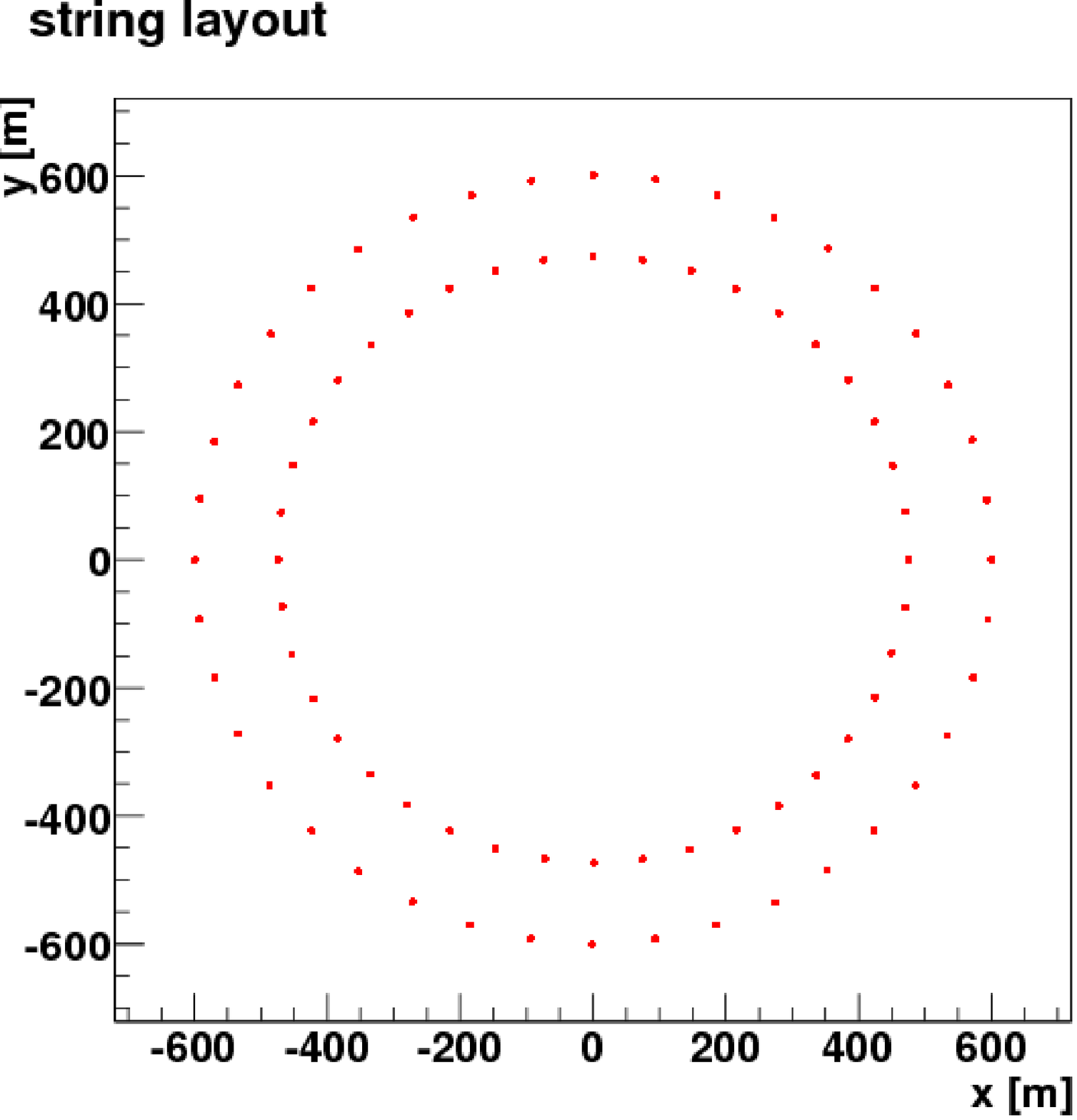}
\label{fig:km3net-ring}}
\hfil
\subfigure[Cluster]{\includegraphics[height=2.in,angle=-90]{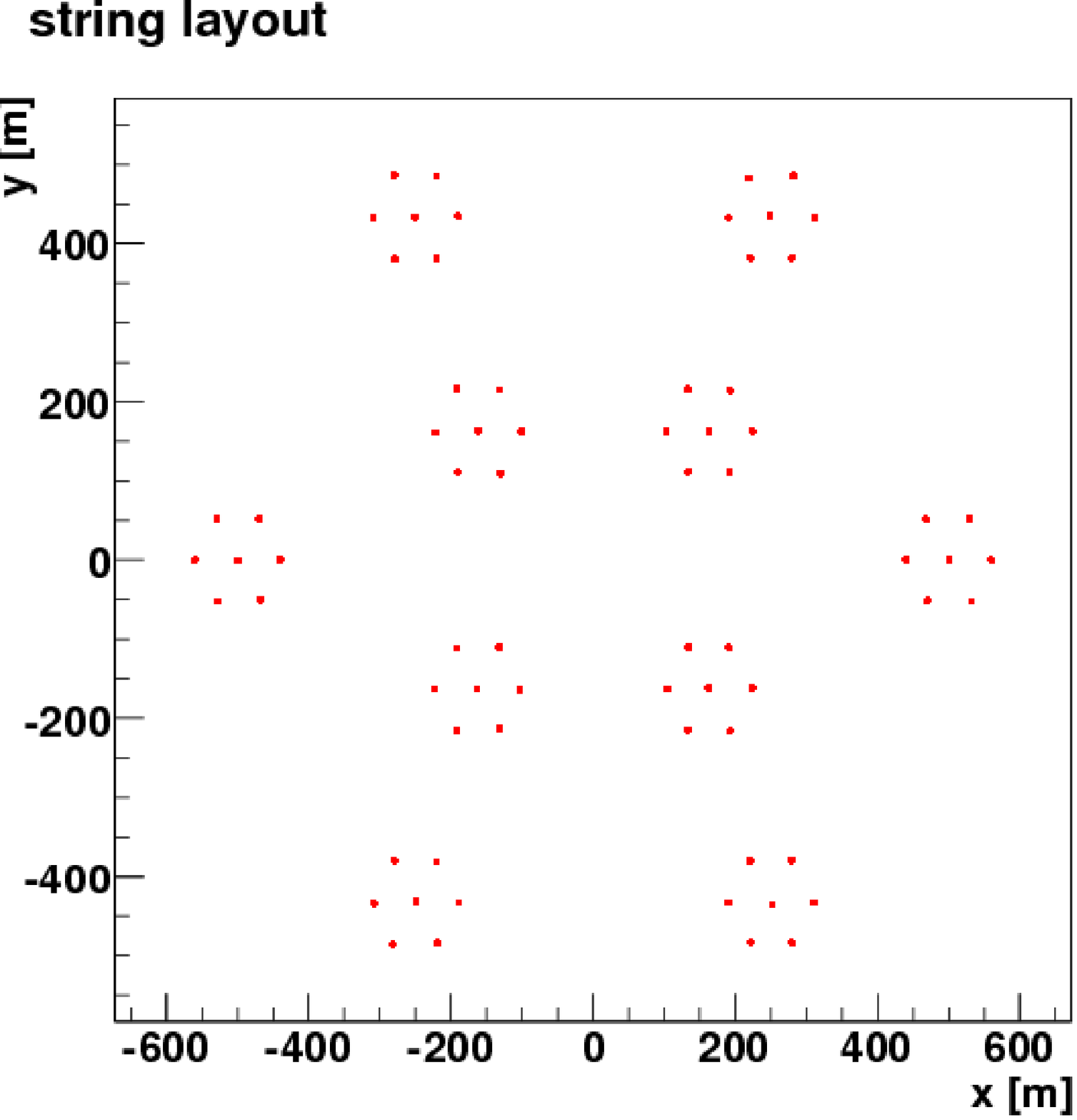}
\label{fig:km3net-}}}
\caption{Examples of different seafloor layouts. Each dot represents a
vertical structure (detector unit) with several OMs (taken from
\cite{thesis:kuch:2007}).}
\label{fig:km3net-detector_configs}
\end{figure*}

The OM configurations were then evaluated in different spatial
arrangements (detector layouts) \cite{thesis:kuch:2007}. The OMs are
positioned with equal spacing in a vertical structure (detector
unit). Several of these units are then combined in different seafloor
layouts. Examples can be seen in
Fig.~\ref{fig:km3net-detector_configs}.
In order to compare the intrinsic sensitivity of different detector
designs, the instrumented volume of all detectors was fixed to $1
\km^3$ and the total cathode area was kept approximately constant. 

\begin{figure}
\centering
\includegraphics[width=3.4in]{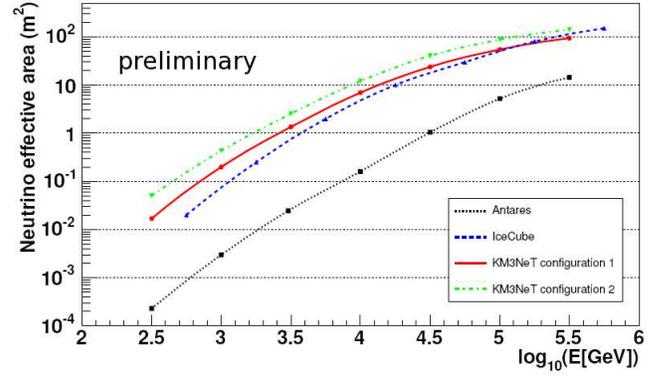}
\caption{Muon effective area as a function of energy for two different
KM3NeT configurations (see main text for a description) and ANTARES
and IceCube.}
\label{fig:km3net-mu_effArea}
\end{figure}

It turns out that a configuration of 225 detector units arranged in a
cuboid grid with an inter-line spacing of $95 \met$, a vertical OM
distance of $16.5 \met$ and 21 3'' PMTs per OM (upper right
configuration in Fig.~\ref{fig:km3net-PMT_layouts}) yields a very good
performance. The resulting muon effective area as function of neutrino
energy is displayed in Fig.~\ref{fig:km3net-mu_effArea} (configuration
2) where it is compared to an alternative configuration~1 (127
detector units with $100 \met$ horizontal spacing; 25 OMs per unit
with $15 \met$ separation; lower right OM configuration in
Fig.~\ref{fig:km3net-PMT_layouts}) as well as IceCube and ANTARES.
Its determination is based on a full simulation of the neutrino
reaction, muon propagation in water and photon detection. Optical
noise equivalent to $40 \khz$ in a 10'' PMT (corresponding to results
from measurements at potential detector sites
\cite{proc:icrc07:chiarusi:1,nim:a552:420}) was also simulated. Event
selection and muon-track reconstruction algorithms were applied.

\begin{figure}
\centering
\includegraphics[width=3.2in]{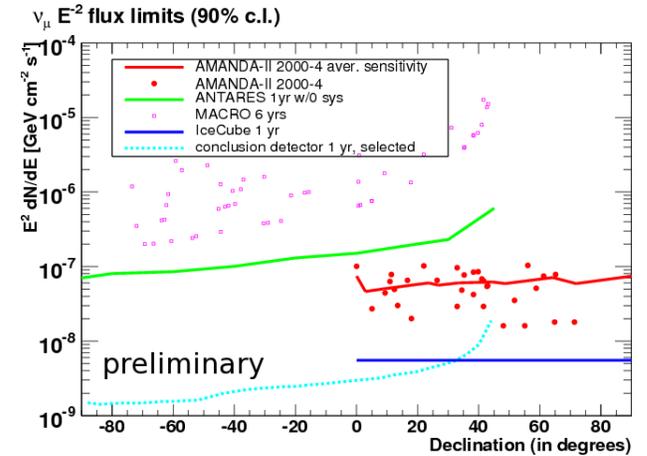}
\caption{Sensitivity of a possible KM3NeT configuration
(blue dashed line labelled \emph{conclusion detector}; equivalent to
configuration 1 in Fig.~\ref{fig:km3net-mu_effArea}) to point-like
sources as a function of declination. Also shown are the sensitivity
of IceCube and other experiments (taken from
\cite{thesis:kuch:2007}).}
\label{fig:km3net-ps_sensitivity}
\end{figure}

The detector configuration~1 was also used to determine the
sensitivity of KM3NeT to point sources. In
Fig.~\ref{fig:km3net-ps_sensitivity} this is compared to the
sensitivity of several other experiments.
The improved sensitivity as compared to IceCube can be largely
explained by the three times larger photo-cathode area and to lesser
extend by the better angular resolution and the non-consideration of
background from miss-reconstructed atmospheric muons. In this
configuration KM3NeT will be able to search the southern sky with a
sensitivity more than 10 times higher than current experiments.

\section{R\&D of detector components}
In overcoming the technical challenges of constructing a neutrino
telescope in the deep sea, the collaboration can build on the
long-lasting experience from a series of neutrino telescopes
experiments both in water (DUMAND, Baikal, ANTARES, NEMO, NESTOR) and
in ice (AMANDA, IceCube). Each of these experiments has created
valuable knowledge on which the design of a large detector in a highly
hostile environment is now based. In the following, examples of
technical challenges and ideas for their solutions are discussed.

A crucial parameter for the successful operation of the detector is
the reliability of all off-shore components. Although, in contrast to
detectors frozen into the ice, it is possible to recover and repair
deep-sea parts of the detector, this requires a large amount of time
and resources. Also, due to the large number of components (more than
200 detector units with roughly 10\,000 OMs), even a moderate failure
rate is unacceptable. Therefore, the reliability of all deep-sea
components has been one of the prime guide lines in the design of the
hardware from the begin of the Design Study.

One strategy to improve the reliability is to simplify the hardware as
much as possible. For example, in ANTARES each OM is connected to an
electronics container that---apart of communicating with the OM---also
contains electronics for the optical and acoustic
calibration. Currently, it is investigated whether the calibration
devices can be completely separated from the photon detection units
and mounted on separate lines communicating with the detection units
by light only. A further reduction of complexity currently under
investigation affects the readout scheme of the PMTs. For KM3NeT the
feasibility of an ``all-data-to-shore'' concept is discussed which
would require the usage of optical fibres. Therefore, in order to
reduce the needed electronics as much as possible, a photonics-based
network is under consideration
\cite{proc:icrc07:kooijman:1}. The electrical signal coming from the
OM is ``imprinted'' with an optical modulator on an optical signal
coming from shore. The modulated signals are then transferred back to
shore where they are timestamped. In this scheme no off-shore laser,
digitisation or timing electronics are required.

\begin{figure*}
\centering
\includegraphics[width=6.8in]{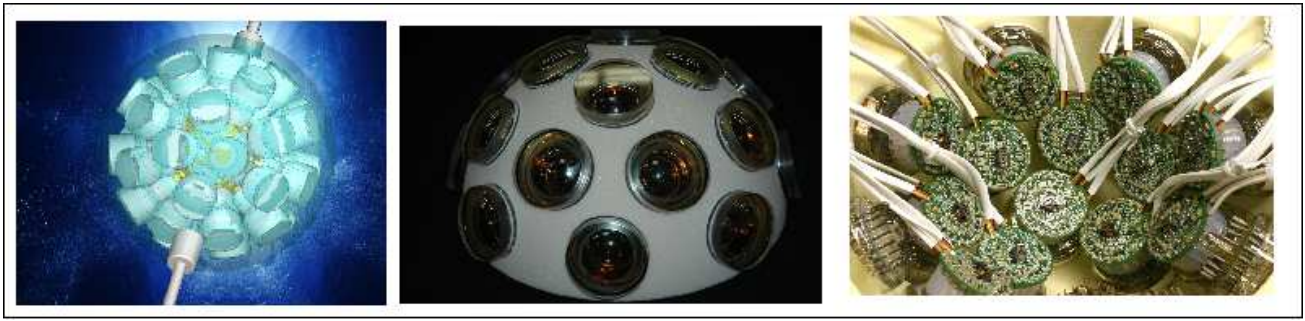}
\caption{Technical drawing (left) and first prototype (middle and
right) of an OM with several small PMTs (taken from
\cite{proc:icrc07:kooijman:2}).}
\label{fig:km3net-multi_PMT_sphere}
\end{figure*}

As shown in Fig.~\ref{fig:km3net-mu_effArea} and discussed in the
previous section OMs equipped with several small PMTs show a very good
performance and several further advantages. A prototype of such an OM
is currently being built and tested at Nikhef
\cite{proc:icrc07:kooijman:2}. Pictures can be seen in
Fig.~\ref{fig:km3net-multi_PMT_sphere}.
As an individual readout is probably not feasible due to the required
bandwidth, the idea would be to reduce the signal from the individual
OMs to a digital pulse with a length equal to the time-over-threshold
of the signal. The rectangular pulses of all PMTs inside an OM are
then superimposed with their proper timing and sent to shore. In this
way sufficient information can be retained afterwards. 

\begin{figure}
\centerline{\subfigure[Segmented PMT]{\includegraphics[height=1.5in]{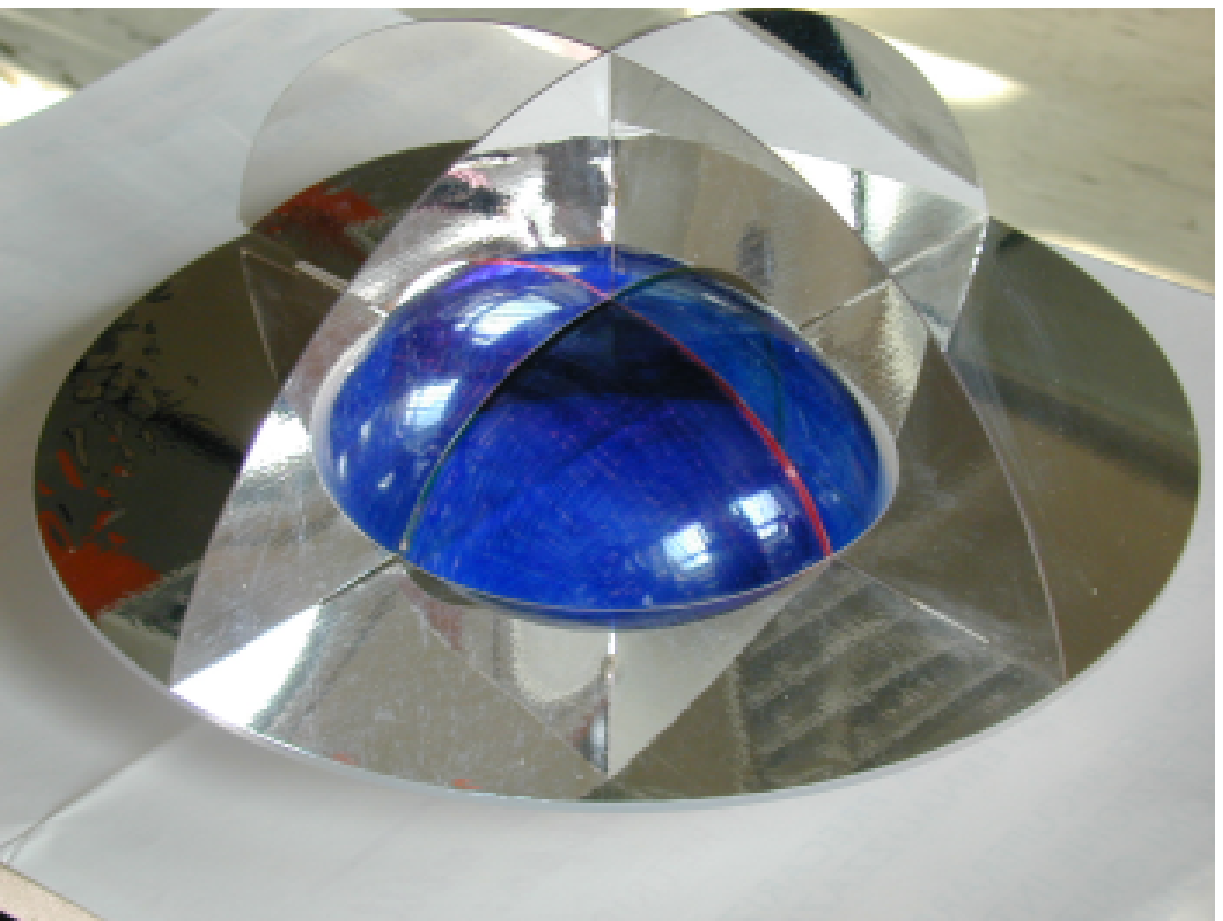}
\label{fig:segmented_pmt}}
\hfil
\subfigure[HPD]{\includegraphics[height=1.5in]{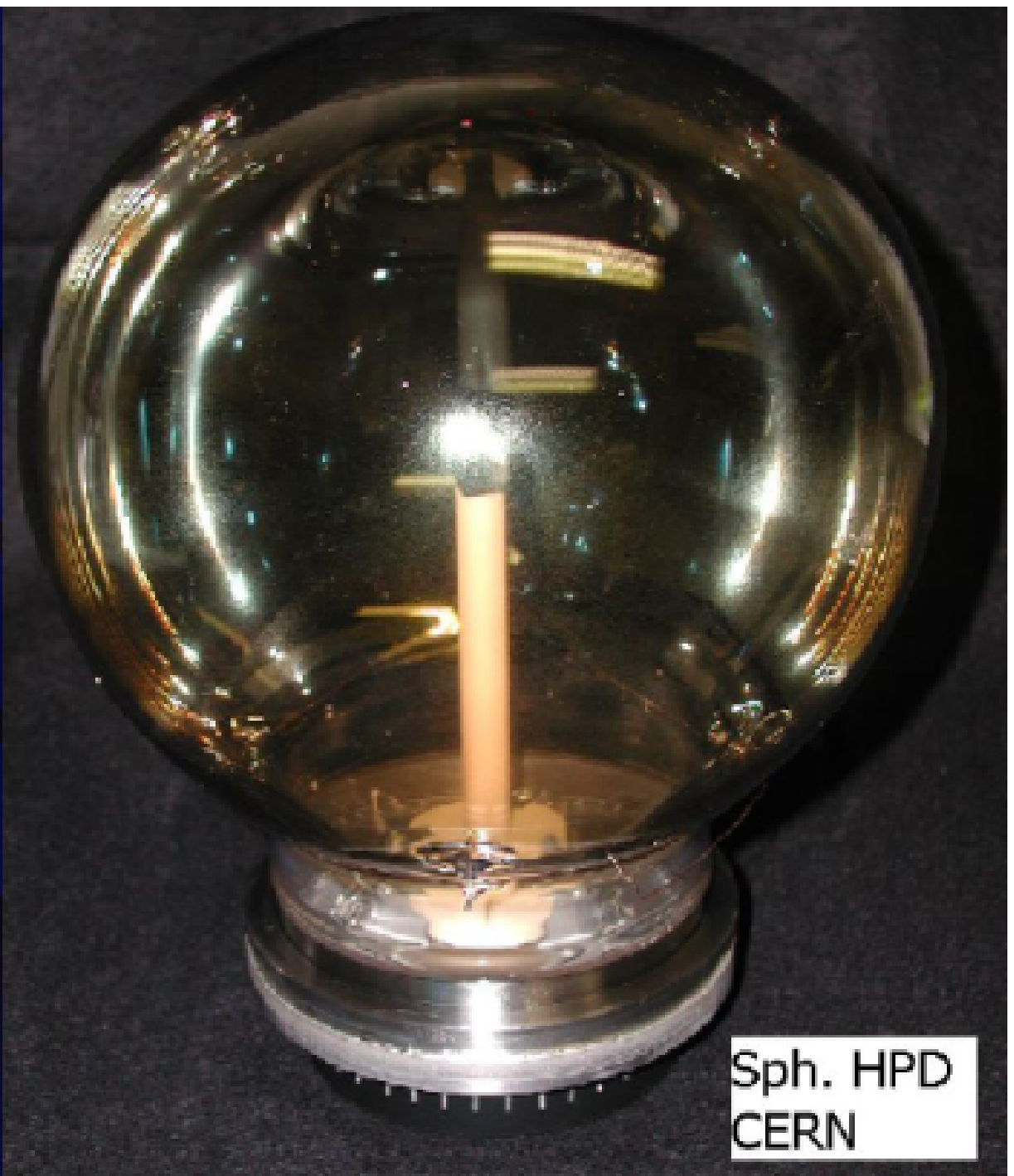}
\label{fig:hpd_pmt}}}
\caption{(a) Prototype of a four-anode 10'' PMT developed by Hamamatsu. (b)
Crystal hybrid photon detector.}
\label{fig:pmt_options}
\end{figure}
Instead of using several small PMTs, another way of obtaining
directional sensitivity currently investigated is the combination of a
segmented with a large multi-anode PMT. A prototype four-anode PMT
developed by Hamamatsu is shown in Fig.~\ref{fig:segmented_pmt}. The
reflecting walls and the floor visible in the picture act like Winston
cones and reflect the photons into the respective quadrant.

Another option being investigated is the usage of crystal hybrid
photon detectors (X-HPDs) (Fig.~\ref{fig:hpd_pmt}). These are large
(up to 15'') tubes of almost spherical geometry in which
photoelectrons emitted from a standard bialkali photocathode are
accelerated in a $\sim\hspace{-2mm}25\,$kV field to bombard a crystal scintillator
viewed by a small PMT. Crystal HPDs have been in operation at the Lake
Baikal detector since 1996. Spherical geometry X-HPDs show a larger
overall efficiency and have a much greater solid angle coverage than
standard large PMTs. Hence, they would allow a larger volume to be
instrumented at approximately the same costs.

Up to now a large cost factor for a deep sea neutrino telescope is the
titanium used for the structures and pressure-resistant components
that get in contact with sea water. Titanium is the only material that
can withstand both the large pressure and the aggressive salt water in
the deep sea. However, a separation of the pressure and corrosion
resistant elements would allow the usage of much cheaper
materials. This idea was realised by the NEMO collaboration in the
design of a junction box
\cite{proc:ecrs:sapienza:1} which consists of four cylindrical
steel vessels hosted in a large, oil-filled fibreglass container. This
avoids the direct contact between steel and sea water. As the KM3NeT
infrastructure will probably require several junction boxes, this may
lead to a significant cost reduction.

In contrast to other (astroparticle) experiments there exists
currently no neutrino source that can be used to independently check
the direction calibration of the neutrino telescope. Though the
direction of a reconstructed track is completely determined by the
position and timing of the photon signals in the PMTs it is highly
desirable to have an independent check for this crucial
parameter. Therefore, the requirements for and feasibility of a
floating surface array somewhat similar to IceTop at the South Pole
are currently investigated. Its position can be determined very
precisely with GPS receivers and it would use atmospheric muons to
calibrate the deep sea detector. According to preliminary results from
these studies three stations at distances of $20 \met$, each consisting
of $16 \met^2$ of scintillating hodoscopes are sufficient. Apart from
the direction calibration, such a surface array would also allow to
verify the efficiency and angular resolution of the detector.

\section{Production model and sea operations}
Not only the design of the components but also the production model
poses a major challenge. A realistic detector configuration consists
of 10\,000 OMs on 250 detector units. For calibration purposes about
25 further dedicated lines are required. The construction should not
take longer than about 3 years in order to take data concurrently with
IceCube over a long time period. This results in the requirement to
produce 15 OMs per day and integrate and test 10 detector and 1
calibration units per month. The latter will probably need 5 assembly
sites and an elaborate logistics to deliver all required components
from the different locations in time.

\begin{figure}
\centering
\includegraphics[width=3.in]{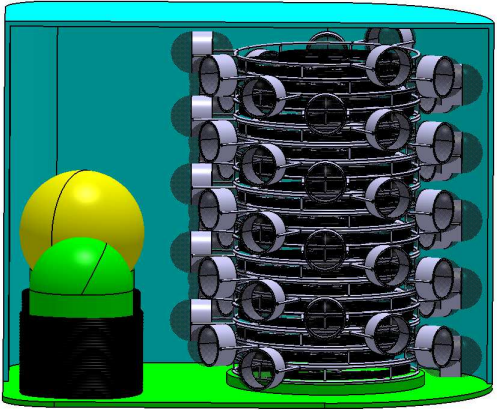}
\caption{Technical drawing of a possible packaging scheme for a
detector unit prior to deployment. The rolled up detector unit is
located on the right side and the buoy (yellow) and release mechanism
(black+green) are depicted on the left side.} 
\label{fig:km3net-sampleline}
\end{figure}

The large number of detector units also implies that new ways of
detector deployment have to be developed. For example, in the case of
ANTARES (12 lines) each line is deployed separately which takes
typically 6 hours. Afterwards, in a separate campaign the lines are
connected to a junction box with a submersible which again takes
several hours.  This scheme is clearly impractical for a telescope
with 250 detector units. A possible solution is the deployment of
``compacted'' detector units similar to the NEMO scheme
\cite{proc:ecrs:sapienza:1}. An example is displayed in
Fig.~\ref{fig:km3net-sampleline}.
Here, the rolled-up detector unit resides in a container together with
the buoy and a release mechanism.  After deployment of the container
on the sea floor the release mechanism is triggered by an acoustic
signal and the buoyancy of the buoy unfolds the detector unit. The
deployment of several containers already interconnected at the surface
reduces the underwater operations and further speeds up the deployment
process. That this is a viable option was demonstrated by the NESTOR
collaboration which deployed their tower module without any underwater
operation
\cite{proc:vlvnt2:anassontzis:1}. Apart from the time saved by
interconnecting the detector units already at the surface this also
reduces the number of wet-mateable connectors which are a considerable
cost factor (each costs several thousand Euro) and are potential
points of failure.

\section{Associated sciences}
Already the continuous measurements of the deep-sea bioluminescence
rate with the ANTARES detector off the cost of Toulon has triggered
much interest in the marine biology community. Permanent deep-sea
installations are rare and most measurements are performed with
battery-powered devices that are deployed for rather short periods of
time. Therefore, the possibility to connect dedicated instruments to a
node with a permanent data and power link to shore is highly
welcome. The importance of this part of the KM3NeT infrastructure is
reflected by a dedicated working group consisting of marine scientists
within the framework of the Design Study.

This so called \emph{Associated Science Node}
\cite{proc:vlvnt:battle:1} will directly branch off the shore cable
and will consist of several junction boxes widely distributed over the
sea floor. Each junction box serves several test sites connected with
tethers which can be moved with remotely operated vehicles
(ROVs). Well defined hard- and software interfaces allow for the
connection of arbitrary observatories.

\section{Conclusions and outlook}
The next generation neutrino telescope in the Mediterranean Sea,
KM3NeT, is well on its way. With the expertise from the construction
of the first generation of deep sea neutrino telescopes an EU funded
Design Study is currently conducted that will provide a TDR by
2009. This TDR will contain the technical specifications for the
future KM3NeT infrastructure consisting of the neutrino telescope and
an associated science node. Subject to the convergence of the
accompanying political process, construction is planned to begin
shortly afterwards and to be finished within 3 years.

The first half of the Design Study has been dedicated to the
exploration of different detector layouts, their physics sensitivity
and technical concepts for their realisation. In preparation of the
TDR a conceptual design report (CDR) will be compiled at the end of
this year containing the results from this first phase. Based on this
information the final detector concept will be elaborated in detail in
the second phase of the Design Study and published in the TDR.

\section*{Acknowledgement}
A. Kappes is currently working at the University of Wis\-consin-Madison,
Department of Physics Madison, Wisconsin 53703, USA. He acknowledges
the support by the EU Marie-Curie OIF program.


\end{document}